# Proximity Full-Text Search with a Response Time Guarantee by Means of Additional Indexes

Alexander B. Veretennikov
Chair of Calculation Mathematics and Computer Science, INSM
Ural Federal University
Yekaterinburg, Russia
alexander@veretennikov.ru

This is a pre-print of a contribution "Veretennikov A.B. Proximity Full-Text Search with a Response Time Guarantee by Means of Additional Indexes" published in "Arai K., Kapoor S., Bhatia R. (eds) Intelligent Systems and Applications. IntelliSys 2018. Advances in Intelligent Systems and Computing, vol 868" published by Springer, Cham. The final authenticated version is available online at: https://doi.org/10.1007/978-3-030-01054-6_66. The work was supported by Act 211 Government of the Russian Federation, contract № 02.A03.21.0006.

*Abstract*—Full-text search engines are important tools for information retrieval. Term proximity is an important factor in relevance score measurement. In a proximity full-text search, we assume that a relevant document contains query terms near each other, especially if the query terms are frequently occurring words. A methodology for high-performance full-text query execution is discussed. We build additional indexes to achieve better efficiency. For a word that occurs in the text, we include in the indexes some information about nearby words. What types of additional indexes do we use? How do we use them? These questions are discussed in this work. We present the results of experiments showing that the average time of search query execution is 44-45 times less than that required when using ordinary inverted indexes.

## I. Introduction

A search query consists of several words. The search result is a list of documents containing these words. In a modern search system, documents that contain search query words that are near each other are more relevant than other documents [1, 2]. Inverted indexes [3, 4] are used to address this search task. For each word in each document, we need to store a record in the index. This record includes the number of the word in the document and the *ID* (identifier) of the document. We can define the *ID* of a document as the document's ordinal number. These records are called "postings".

Words appear in documents at different frequencies. The maximum query response time is determined by the most frequently occurring words. Zipf's law [5] describes the typical word frequency distribution. It is common to have a search system that can usually perform a query within 1 sec. of time but works very slowly, requiring 20-30 sec., for example, for a query that contains frequently occurring words. This is the problem that we wish to solve in this paper. With additional indexes, we can guarantee a stable query response time, for example, within 1 sec.

An example of a word frequency distribution is shown in Fig. 1. The horizontal axis represents different words, from frequently occurring words to infrequently occurring words (from left to right, in decreasing order of their frequencies). On the vertical axis, we plot the number of occurrences of each word. With typical inverted indexes, the query execution time is proportional to the number of occurrences of the queried words in the indexed texts.

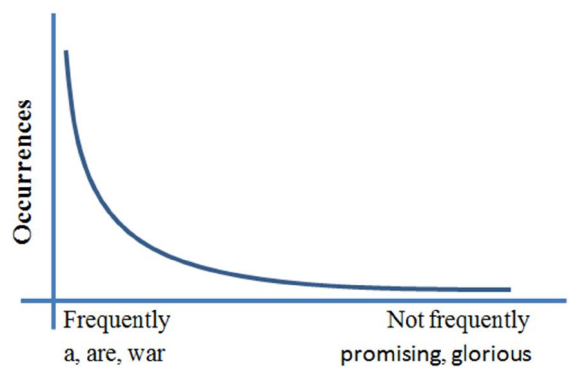

Fig. 1. Word frequency distribution.

Some search systems exclude most frequently used words from the index and, consequently, from any search – this is called the stop words approach. However, this approach is not correct [6]. Some most frequently occurring words can have unique meanings in specific contexts. For example, consider the query "time and a word yes". Yes are an English rock band, and "Time and a Word" is one of their well-known songs. Therefore, the word "yes" has a specific meaning in the context of this query. A similar query example is "who are you who".

The Who are an English rock band, and "Who are You" is one of their songs.

The other approach is to maintain additional indexes for faster query execution.

In [1], some additional indexes are introduced, but only two-word queries are considered. In [7], most frequently occurring words are excluded from consideration. We avoid both of these faults by introducing several types of additional indexes for several types of words.

## II. PROXIMITY AND RELEVANCE

### A. Importance of proximity

We introduce a parameter *MaxDistance*. Let us consider a search result – a document containing words in the specified query. If the length of the fragment containing the words is less than or equal to *MaxDistance*, then the search result is relevant and important; otherwise, it can be skipped.

For example, we consider the search query "time and a word yes".

Result 1: "time and a word yes" – This is an important search result.

Result 2: "time and a word by yes" – This search result is also important.

Result 3: "time …some other words … and … some other words … a … word … some other words… by … some other words… yes" – This search result may not be important because "time", "and", and "yes" are not linked by any meaning.

We will present a formal definition of *MaxDistance* later.

### B. Relevance function

Let the importance of a pair of word occurrences be inversely proportional to the square of the distance between the words in the document [1].

Consider the following relevance function:

$$S = a \cdot SR + b \cdot IR + c \cdot TP \qquad (1)$$

Here $a$, $b$, and $c$ are non-negative parameters, where $a + b + c = 1$. Search results can be sorted in accordance with this function [1]. *SR* is a static rank of the document, such as the PageRank [8], and is independent of the query. *IR* is an information retrieval rank, such as BM25. *TP* is a proximity ranking function.

For two word occurrences $A$ and $B$, the value of *TP(A, B)* can be calculated as

$1 / |A - B|^2$, where $(|A - B|)$ is the distance between the words in the document.

We define the occurrence of a word as its ordinal number within the document.

Consider the following text: "time and a word by yes".

The word positions (ordinal numbers) are as follows: time : 0, and : 1, a : 2, word : 3, by : 4, yes : 5.

Consider the following search query: "and word". The search result is "and a word", at position 1. There is an extra word, "a", between "and" and "word" in this text. Therefore,

$TP = 1 / |1 - 3|^2 = 0.25$.

Consider the search query "time and". The search result is "time and", at position 0. Therefore,

$TP = 1 / |0 - 1|^2 = 1$.

If the search query occurs in the text in its exact form (with no extra words between the query terms in the text), then

$TP = 1$.

### C. Importance of TP

Let us introduce a value denoted by *MaxTPDistance* to define the following condition: if $|A - B| \leq$ *MaxTPDistance*, then *TP* is "important"; otherwise, only (*SR* + *IR*) is "important".

What does "important" mean? We can assume that *SR*, *IR* and *TP* can be normalized.

Thus, $0 \leq SR \leq 1$, $0 \leq IR \leq 1$, and $0 \leq TP \leq 1$.

Then, $0 \leq a \cdot SR \leq 1$, $0 \leq b \cdot IR \leq 1$, and $0 \leq c \cdot TP \leq 1$.

Let us introduce a parameter *TP_Critical*, with an example value of 0.15.

If $(c \cdot TP) \leq$ *TP_Critical*, then the relevance (or score) of the search result will be determined mostly by *SR* + *IR*. In this case, *TP* is not "important".

Let $c = 1$ and *TP_Critical* = 0.15.

Consider the following text: "time and a word by yes".

Consider the following search query: "and word". The search result is "and a word", at position 1, with $|A - B| = 2$.

$c \cdot TP = 0.25 >$ *TP_Critical*, indicating that the search result is important.

Now, consider the search query "time word". The search result is "time and a word", at position 0, with $|A - B| = 3$.

$c \cdot TP = 1 / 3^2 \approx 0.11 <$ *TP_Critical*, indicating that the search result is not important.

In this case, *MaxTPDistance* = 2.

### D. Evaluating TP for more than two words

How do we evaluate *TP* if the query consists of more than two words? Consider an *n*-word query *Q*. We have a search result *R*, which is represented by *n* word positions in a document:

$R = X(1), X(2), …, X(n)$.

Let $A(R) = min(X(1), …, X(n))$ and

$B(R) = max(X(1), …, X(n))$.

Requirement: if the search query occurs in the text in its exact form (with no extra words between the query terms in the text), then $TP(R) = 1$.

$TP(R) = 1$ when $|A(R) - B(R)| = (n - 1)$.

Proposition:

$TP(R) = TP(X(1),…,X(n)) = 1 / (|A(R) - B(R)| - (n - 2))^2$.

For example, suppose that we have the following search query: "time and a word yes".

Result 1: "time and a word yes" – This is an important search result.

$TP(R) = 1 / (|0 - 4| - (5 - 2))^2 = 1$.

Result 2: "time and a word by yes" – This is a less important search result.

$TP(R) = 1 / (|0 - 5| - (5 - 2))^2 = 0.25$.

We also can consider a more flexible *TP* function, such as

$TP(R) = TP(X(1),…,X(n)) =$

$1 / (p \cdot (|A(R) - B(R)| - (n - 2)))^2$.

The value of *p* can be different for different systems.

*E. Evaluating MaxTPDistance for more than two words*

Let us define the function *MaxTPDistance*(*n*) as follows: for any search query *Q* consisting of *m* words, where $m \le n$, and a search result $R = X(1), X(2), …, X(m)$ for *Q*, if $|A(R) - B(R)| > MaxTPDistance(n)$, then $c \cdot TP(R) \le TP\_Critical$; moreover, *MaxTPDistance*(*n*) is the smallest value for which this is true.

By definition,

if $a \ge b$, then *MaxTPDistance*(*a*) $\ge$ *MaxTPDistance*(*b*).

Let $n = 3$, $TP\_Critical = 0.15$, and $c = 1$.

Consider a 3-word search query *Q* and a search result *R*.

If $|A(R) - B(R)| = 2$, then

$TP(R) = 1 / (2 - 1)^2 = 1 > TP\_Critical$.

If $|A(R) - B(R)| = 3$, then

$TP(R) = 1 / (3 - 1)^2 = 0.25 > TP\_Critical$.

If $|A(R) - B(R)| = 4$, then

$TP(R) = 1 / (4 - 1)^2 \approx 0.11 < TP\_Critical$.

Consider a 2-word search query *Q* and a search result *R*.

If $|A(R) - B(R)| = 1$, then

$TP(R) = 1 / (1)^2 = 1 > TP\_Critical$.

If $|A(R) - B(R)| = 2$, then

$TP(R) = 1 / (2)^2 = 0.25 > TP\_Critical$.

If $|A(R) - B(R)| = 3$, then

$TP(R) = 1 / (3)^2 \approx 0.11 < TP\_Critical$.

In this case, *MaxTPDistance*(3) = 3.

For any query *Q* consisting of *m* words, where $m \le 3$, and any search result *R* for *Q* that satisfies the condition $|A(R) - B(R)| > 3$, we have $c \cdot TP(R) \le TP\_Critical$.

*F. Definition of MaxDistance*

Let us introduce our new parameter, *MaxDistance*.

Let $n \ge 1$ be a number.

We assume that for any query of length $m \le n$, our search will return all relevant results. If the query has a length $> n$, it must be divided into parts.

Let *MaxDistance* = *MaxTPDistance*(*n*).

We can also define a parameter *MaxDistance* = 7 (for example) and build indexes accordingly. Then, for any query of length *m*, where $m \le n \le MaxDistance$, with *n* being some number, our search will return all relevant results.

In our experiments, we use *MaxDistance* = 5, 7 or 9.

*G. More generic TP structure.*

Let us also consider a more generic version of *TP*:

$TP(R) = TP(X(1),…,X(n)) = 1 / (|A(R) - B(R)| - (n - 2))^{e(n)}$,

$e(n) = 1 + (2 / n)$.

We assume that for longer queries, more extra words are acceptable between query terms in the text.

Let us calculate *MaxTPDistance*(3) for this case.

Let $n = 3$, $TP\_Critical = 0.15$, and $c = 1$.

Consider a 3-word search query *Q* and a search result *R*.

If $|A(R) - B(R)| = 2$, then $TP(R) = 1 > TP\_Critical$.

If $|A(R) - B(R)| = 3$, then $TP(R) \approx 0.314 > TP\_Critical$.

If $|A(R) - B(R)| = 4$, then $TP(R) \approx 0.16 > TP\_Critical$.

If $|A(R) - B(R)| = 5$, then $TP(R) \approx 0.09 < TP\_Critical$.

Consider a 2-word search query *Q* and a search result *R*.

If $|A(R) - B(R)| = 1$, then

$TP(R) = 1 / (1)^2 = 1 > TP\_Critical$.

If $|A(R) - B(R)| = 2$, then

$TP(R) = 1 / (2)^2 = 0.25 > TP\_Critical$.

If $|A(R) - B(R)| = 3$, then

$TP(R) = 1 / (3)^2 \approx 0.11 < TP\_Critical$.

In this case, *MaxTPDistance*(3) = 4. We need a larger value of *MaxDistance* with such a *TP* function.

### III. WORD TYPE

In [9], we defined three types of words.

**Stop words:** Examples include "and", "at", "or", "yes", "who", "was", and "war". These words are very commonly encountered and may not be included in the index in some other approaches. However, we include all words.

**Frequently used words:** These words are frequently encountered but convey meaning. These words should always be included in the index.

**Ordinary words**: This category contains all other words. We assume that no performance problems will arise from these words.

We use a morphological analyzer for lemmatization. For each word in the dictionary, the analyzer provides a list of numbers of lemmas (i. e., basic or canonical forms). The lemma numbers lie in the range from zero to (*WordsCount* – 1), where *WordsCount* is the number of different lemmas considered (we use a combined Russian/English dictionary with approximately 200 000 Russian lemmas and 92 000 English lemmas).

If a word does not appear in the analyzer's dictionary, we assume that its lemma is the same as the word itself.

When using the analyzer, we apply aforementioned three-type division approach, not to the words themselves but to the lemmas of the words. The lemmas are divided into three types in terms of the frequency with which they are encountered: stop lemmas, frequently used lemmas, and other lemmas.

How do we distribute the lemmas among these groups? Let us sort all lemmas in decreasing order of their occurrence frequency in the texts. This sorted list we call the *FL*-list. The number of a lemma in the *FL*-list we call its *FL*-number. Let the *FL*-number of a lemma $w$ be denoted by $FL(w)$.

The first *SWCount* most frequently occurring lemmas are stop lemmas. The second *FUCount* most frequently occurring lemmas are *frequently used* lemmas. All other lemmas are ordinary lemmas. *SWCount* and *FUCount* are parameters. Representative example values are *SWCount* = 700 and *FUCount* = 2100.

Let us consider the following text, with identifier *ID*1: "A friend of mine who has desired the honour of meeting with you". This is the excerpt from the Charles Dickens's Barnaby Rudge.

After lemmatization: [a] [friend] [of] [mine, my] [who] [have] [desire] [the] [honour] [of] [meet, meeting] [with] [you].

With *FL*-numbers: [a: 17] [friend: 793] [of: 24] [mine: 2482, my: 264] [who: 293] [have: 55] [desire: 2163] [the: 10] [honour: 3774] [of: 24] [meet: 1008, meeting: 4375] [with: 40] [you: 47].

Let us enumerate the words starting from zero. Then, the word "friend" appears in the text at position 1. Then, the lemma "friend" appears in the text at position 1. The lemma "my" appears in the text at position 3. Thus, the distance between the lemma "my" and the lemma "friend" in the text is 2. We can say that lemma "my" > "of", because $FL(my) = 264$, $FL(of) = 24$, and $264 > 24$ (we use the *FL*-numbers to establish the order of the lemmas in the set of all lemmas).

For an ordinary lemma $q$, we can say that $FL(q) = \sim$. In this case, $q$ occurs in the texts so rarely that $FL(q)$ is irrelevant. We denote by "~" some big number.

Let us consider the results obtained with our example values, namely, *SWCount* = 700 and *FUCount* = 2100.

Stop lemmas (< 700): a, of, my, who, have, the, with, you.

Frequently used lemmas ($\geq$ 700, < 2800): friend, mine, desire, meet.

Ordinary lemmas ($\geq$ 2800): honour, meeting.

## IV. Additional indexes

We define several types of additional indexes.

### A. The ordinary index with near stop word (NSW) records

For each lemma in each document, a record (*ID*, *P*, *NSW*) is included in the index. *ID* is the ordinal number of the document. *P* is the corresponding word's ordinal number within the document. The NSW record contains information about all stop lemmas occurring near position *P* (at a distance ≤ *MaxDistance*). This information is efficiently encoded [9, 10, 11].

For example, let *MaxDistance* = 5. The NSW record for the first occurrence of "friend" in the aforementioned example contains the following: (a, −1), (of, 1), (my, 2), (who, 3) (have, 4). In (a, −1), the distance (−1) between "a" and "friend" is stored, and so on.

Let the document identifier *ID*1 be 27. Below, let us consider several example postings in the ordinary index, in the format (document *ID*, word position, NSW record).

friend: (27, 1, ((a, −1), (of, 1), (my, 2), (who, 3) (have, 4))).

mine: (27, 3, ((a, −3), (of, −1), (who, 1), (have, 2), (the, 4))).

desire: (27, 6, ((of, −4), (my, −3), (who, −2), (have, −1), (the, 1), (of, 3), (with, 5))).

The lemma types considered here are frequently used and ordinary.

For a stop lemma, we include only the first occurrence in the document and no NSW records.

Let us consider a lemma. For optimization purposes, we can use two data streams for the lemma. The first data stream contains the (*ID*, *P*) records. The second data stream contains the corresponding NSW records. In this case, the NSW records can be easily skipped if required.

See [11] for more details about NSW records.

### B. The expanded (w, v) indexes.

The expanded ($w$, $v$) index is the list of occurrences of the lemma $w$ for which lemma $v$ occurs in the text at a distance less than or equal to *MaxDistance* from $w$.

The lemma types considered are as follows: for $w$, frequently used; for $v$, frequently used or ordinary. Each posting includes the distance between $w$ and $v$ in the text.

In the case that both $w$ and $v$ are frequently used, we create only one expanded index. To prevent duplication, we create an expanded ($w$, $v$) index only if $w \leq v$.

Below, let us consider several example postings in the format (document *ID*, word position, distance).

(friend, mine): (27, 1, 2). Here, 27 is the document *ID*, 1 is the position of "friend" in the document, and 2 is the distance between "friend" and "mine" in the document. We store a posting only for the key (friend, mine); no posting is stored for the key (mine, friend). Note that *FL*(friend) = 792 < *FL*(mine) = 2482.

(friend, desire): (27, 1, 5).

(desire, mine): (27, 6, −3).

(mine, honour): (27, 3, 5).

We previously studied (*w*, *v*) indexes in [9, 10, 11, 12]. In the current work, we describe several new use cases.

*C. The expanded (f, s, t) indexes*

The expanded (*f*, *s*, *t*) index is the list of occurrences of the lemma *f* for which lemmas *s* and *t* both occur in the text at distances less than or equal to *MaxDistance* from *f*.

We create an expanded (*f*, *s*, *t*) index only for the case in which $f \leq s \leq t$.

Here, *f*, *s*, and *t* are all stop lemmas.

Below, let us consider several example postings, in the format (document *ID*, word position, distance between *f* and *s*, distance between *f* and *t*):

(a, of, my): (27, 0, 2, 3).

(a, my, who): (27, 0, 3, 4).

(a, of, who): (27, 0, 2, 4).

(a, have, my): (27, 0, 5, 3).

(of, my, who): (27, 2, 1, 2).

…

(of, with, who): (27, 9, 2, −5).

This type of index is the largest.

## V.   PREPROCESSING THE QUERY

Let us consider the following query: "friend mine who".

After lemmatization: [friend] [mine, my] [who].

Each element of the query after lemmatization is called a cell. This query contains three cells. The first cell is [friend], the second is [mine, my], and the last is [who].

Important condition: Each cell of the query must contain lemmas of only one type. If this condition is not met, then the query must be divided. For example, from the initial query [friend] [mine, my] [who], we derive two queries: [friend] [mine] [who] and [friend] [my] [who].

Second condition: If all lemmas in the query are stop lemmas, then each cell must contain only one lemma. If this condition is not met, then the query must be divided.

## VI.   PROCESSING THE QUERY

We apply different processing methods for different types of queries.

*A. All lemmas of the query are ordinary.*

In this case, we use the ordinary index. We skip the NSW records. In this case, *MaxDistance* is not used.

*B. All lemmas of the query are frequently used*

Let us consider the following query: "beautiful red hair".

After lemmatization: [beautiful: 2216] [red: 2191] [hair: 1850].

*1) The first approach*

The lemma "beautiful" is the lemma that is encountered *least often* in the texts.

We therefore designate [beautiful] as the main cell of the query.

We consider the following expanded indexes: (red, beautiful) and (hair, beautiful).

The lemma "red" is more frequently used than "beautiful". Thus, we have the expanded index (red, beautiful). However, there are two logical expanded indexes: (red, beautiful) and (beautiful, red). Suppose that we are reading records from the (red, beautiful) index. From a record (*ID*, *Position*, *Distance*), we can produce the record (*ID*, *Position* + *Distance*, −*Distance*), which corresponds to the (beautiful, red) logical index.

Thus, we can say that we have the (beautiful, red) and (beautiful, hair) indexes.

Each of these indexes contains the positions of the lemma "beautiful" in texts.

Let us consider the following text: "A beautiful, shimmering, red curly hair …".

We have (*ID*, 3, –2) in the (red, beautiful) index and (*ID*, 5, –4) in the (hair, beautiful) index.

Consequently, we have (*ID*, 1, 2) in the (beautiful, red) logical index and (*ID*, 1, 4) in the (beautiful, hair) logical index. These two records have identical *ID* and *Position* fields.

We need to check for all lemmas in the query, except the main lemma, for which a record (*ID*, *P*, *) exists at the specified position (*ID*, *P*) in all selected logical expanded indexes.

Let us consider the position (*ID*, 1). In the (beautiful, red) index, a record (*ID*, 1, 2) exists. In the (beautiful, hair) index, a record (*ID*, 1, 4) exists. Thus, we have "beautiful red hair" in the text.

*2) The second approach.*

Let us consider a (*w*, *v*) index. From a record (*ID*, *Position*, *Distance*), we can produce two related records: (*ID*, *Position*) for the key *w* and (*ID*, *Position* + *Distance*) for the key *v*. Thus, from one (*w*, *v*) index, we can derive two logical indexes (*w*, *v*) and (*v*, *w*). The first contains the occurrences of *w*, and the second contains the occurrences of *v*.

From the record (*ID*, 3, –2) in the (red, beautiful) index, we can produce the record (*ID*, 3) as an occurrence of "red" and (ID, 1) as an occurrence of "beautiful".

We can divide any query into a list of pairs of words. For example, let us consider "beautiful red hair" –> (beautiful red) (red hair). Then, from the extended (beautiful, red) and (red, hair) indexes, we can produce 4 streams of postings, one each for "beautiful", "red", "red" and "hair" (there is one extra "red" stream, which we can skip). Then, we combine these logical streams for "beautiful", "red" and "hair" and obtain the results.

*3) Comparison of the first and second approaches*

The second approach requires more computational resources to produce single-key streams in memory, but fewer data need to be loaded. Consider the query "beautiful bright red hair" –> [beautiful: 2216] [bright: 2530] [red: 2191] [hair: 1850].

The first approach requires 3 two-key indexes: (beautiful, bright), (red, bright), and (hair, bright). The second approach requires 2 two-key indexes: (beautiful, bright) and (red, hair).

Now, let us consider the query "beautiful red rose" –> [beautiful: 2216] [red: 2191] [rose: 1007, rise: 1753]. Using the first approach, we need three indexes: (red, beautiful), (rise, beautiful), and (rose, beautiful).

*4) The third approach.*

We can divide any query into a list of pairs of words. For example, let us consider "beautiful red hair" –> (beautiful red) (red hair). Then, we need to combine the corresponding streams of data. This approach is more effective than the second approach, but it is also more complex to realize because it is more complex to combine two-key streams than single-key streams.

C. *Not all of the lemmas are frequently used, and there are no stop lemmas.*

Let us consider the following query: "red glorious promising rose".

After lemmatization: [red: 2191], [glorious: ~] [promising: ~] [rose: 1007, rise: 1753].

Frequently used lemmas: red, rose, rise.

Ordinary lemmas: glorious, promising.

There are several approaches we can propose here.

*1) The first approach.*

We select the frequently used lemma *w* in the query that has the lowest frequency. For every other lemma *v* in the query, a logical expanded (*w*, *v*) index exists. For example, let us select [red] as the main cell. We can use the following expanded indexes:

(red, promising) – contains occurrences of red (near promising).

(red, glorious) – contains occurrences of red (near glorious).

(red, rise) – contains occurrences of red (near rise).

(red, rose) – contains occurrences of red (near rose).

*2) The second approach*

We select the ordinary lemma *w* in the query that has the lowest frequency. For every other frequently used lemma *v* in the query, a logical expanded (*w*, *v*) index exists. For every other ordinary lemma *q* in the query, we can use the ordinary index *q* (skipping the NSW records). For example, let us select [promising] as the main cell. We can use the following indexes:

(red, promising) – contains occurrences of red (near promising).

(rise, promising) – contains occurrences of rise (near promising).

(rose, promising) – contains occurrences of rose (near promising).

(glorious) – we use the ordinary index, because both "glorious" and "promising" are ordinary lemmas and no extended (promising, glorious) index exists.

We do not need a list of occurrences of "promising" because we know that "promising" occurs somewhere nearby.

*3) The third approach.*

We can also select a two-component key index for a frequently used or ordinary lemma. For this example, we have

(red, promising) for red,

(rise, promising) for rise,

(rose, promising) for rose, and

(red, glorious) for glorious.

If we store in some dictionary the length of each index (*w*, *v*), then we can select the most suitable variant.

D. *All lemmas of the query are stop lemmas.*

In this case, (*f*, *s*, *t*) indexes are used.

Let us consider the following query: "to be not to be".

After lemmatization: [to: 7] [be: 21] [not: 156] [to: 7] [be: 21]. We can use the (to, be, not) and (to, to, be) indexes to produce results.

Now, let us consider the following query: "who are you who".

[who: 293] [are: 268, be: 21] [you: 47] [who, 293].

We produce two new queries:

Q1: [who: 293] [are: 268], [you: 47] [who, 293].

Q2: [who: 293] [be: 21], [you: 47] [who, 293].

Let us consider Q1.

We can use the (you, are, who) and (you, who, who) indexes to obtain results.

E. *All lemma types appear in the query*

Let us consider the following query: "notes about Gallic war".

After lemmatization: [note: 1373] [about: 211] [gallic: ~] [war: 674].

Stop lemmas: about, war.

Frequently used lemmas: note.

Ordinary lemmas: gallic.

We select the non-stop lemma *w* with the lowest frequency. For the lemma *w*, we use the ordinary index and process the NSW records. For this example, we select "gallic".

For every other frequently used lemma *v* in the query, a logical expanded (*w*, *v*) index exists. In this example, the only index of this type is (note, gallic).

For every other ordinary lemma *q* in the query, we need to use the ordinary index *q* (skipping the NSW records). If another frequently used lemma *p* exists in the query, we can also use the expanded (*p*, *q*) index instead of the ordinary index.

*F. Additional examples*

Consider the following query: "time and a word yes".

After lemmatization: [time: 184] [and: 28] [a: 17] [word: 602] [yes: 2375].

We can see that in our dictionary, "time", "and", "a", and "word" are all stop lemmas, whereas "yes" is a frequently used lemma. In this case, we can use the ordinary index with NSW records. We select from the ordinary index all occurrences of "yes", and for each such occurrence, we need to check the NSW record for the existence of the lemmas "time", "and", "a", and "word".

VII. SEARCH EXPERIMENT ENVIRONMENT

All search experiments were conducted using a collection of texts with a total size of 71.5 GB, consisting of 195 000 documents of plain text, fiction and magazine articles.

*MaxDistance* = 5, 7 or 9.

*SWCount* = 700.

*FUCount* = 2100.

We used the following computational resources:

CPU: Intel(R) Core(TM) i7 CPU 920 @ 2.67 GHz.

HDD: 7200 RPM.

RAM: 24 GB.

OS: Microsoft Windows 2008 R2 Enterprise.

Query selection: We selected a document from the collection. Next, we selected some words from the document. We formed a query from those words. We selected the words from different positions in the document. We evaluated the query using standard inverted indexes and our indexes to estimate the performance gain of our approach.

The experimental procedure is as follows.

1. Selection of a random document in the index.

2. Selection of search queries as follows.

2.1. Selection of a sequence of words. The query length is 3, 4 or 5.

2.2. Selection of a sequence of words, with the omission of every other word. The query length is 3.

Let us consider a document "Gaul, taken as a whole, is divided into three parts". We select queries "Gaul taken as", "Gaul taken as a", "Gaul taken as a whole" at 2.1. We select "Gaul as whole" at 2.2.

2.3. Selection of a sequence of words, with the omission of the second word. For example, consider the query "Gaul as a whole". The query length is 3 or 4.

2.4. Selection of a sequence of words, with the omission of the second and third word. For example, consider the query "Gaul a whole". The query length is 3.

3. Search for each selected query. We evaluate the query using standard inverted indexes and our indexes. In the search, all the records corresponding to the given word are read. Thus, even if the required query is found, reading continues to the end.

Queries of three, four, or five words are selected, because *MaxDistance* = 5 in the first experiment.

However, we can perform larger queries with a larger value of *MaxDistance*.

The benefits of this approach are as follows.

1. We verify that the index is correctly constructed and performs as required. Since queries are selected from an already-indexed document, they should be precisely found. We verify that the search results include a record corresponding to the document used in selecting the query.

2. The queries found are relatively diverse and include a large number of different words.

3. Many of the queries include stop words and frequently encountered words.

All the queries are processed sequentially in a single program thread.

VIII. SEARCH EXPERIMENTS WITH *MAXDISTANCE* = 5

Idx1: ordinary inverted file without any improvements such as NSW records (the size of Idx1 was 43.3 GB).

Idx2: our indexes, including the ordinary inverted index with NSW records and the (*w*, *v*) and (*f*, *s*, *t*) indexes, with *MaxDistance* = 5.

Queries: 5250 (519 queries consisted only of stop lemmas).

Query length: from 3 to 5 words.

Average query times:

Idx1: 13.66 sec., Idx2: 0.29 sec.

Average data read sizes per query:

Idx1: 468.6 MB, Idx2: 9.9 MB.

We improved the query processing time by a factor of 47.1 with Idx2, and we improved the data read size by a factor of 47.3; see Fig. 2 and Fig. 3, respectively.

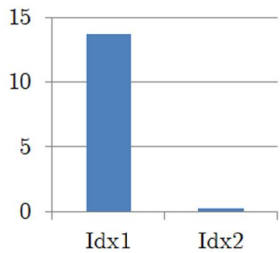

Fig. 2. Average query execution times for Idx1 and Idx2 (seconds), with MaxDistance = 5.

The left-hand bar shows the average query execution time with the standard inverted indexes. The right-hand bar shows the average query execution time with our indexes. Our bar is much smaller than the other bar because our searches are very quick.

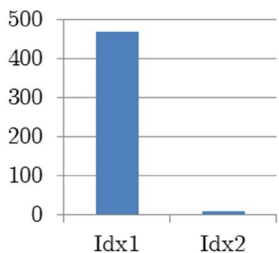

Fig. 3. Average data read sizes per query for Idx1 and Idx2 (MB), with MaxDistance = 5.

The left-hand bar shows the average data read size per query with the standard inverted indexes. The right-hand bar shows the average data read size per query with our indexes. We need to read much fewer data from the disk, and our bar is much smaller than the other bar.

Index sizes:

Ordinary index with NSW records: 110 GB (the total size of the NSW records can be calculated as follows: 110 GB – 43.3 GB = 66.7 GB).

Expanded ($w$, $v$) indexes: 143 GB.

Expanded ($f$, $s$, $t$) indexes: 622 GB.

IX. SEARCH EXPERIMENTS WITH MAXDISTANCE = 7

Idx1: ordinary inverted file without any improvements such as NSW records.

Idx2: our indexes, including the ordinary inverted index with NSW records and the ($w$, $v$) and ($f$, $s$, $t$) indexes, with *MaxDistance* = 7.

Queries: 5250 (519 queries consisted only of stop lemmas).

Query length: from 3 to 5 words.

Average query times:

Idx1: 13.66 sec., Idx2: 0.31 sec.

Average data read sizes per query:

Idx1: 468.6 MB, Idx2: 10.03 MB.

We improved the query processing time by a factor of 44 with Idx2, and we improved the data read size by a factor of 46.7.

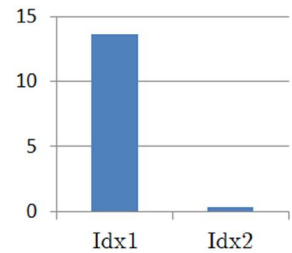

Fig. 4. Average query execution times for Idx1 and Idx2 (seconds), with MaxDistance = 7.

We can see a small increase in the average query execution time in comparison with the *MaxDistance* = 5 case.

X. SEARCH EXPERIMENTS WITH MAXDISTANCE = 9

Idx1: ordinary inverted file without any improvements such as NSW records.

Idx2: our indexes, including the ordinary inverted index with NSW records and the ($w$, $v$) and ($f$, $s$, $t$) indexes, with *MaxDistance* = 9.

Queries: 5250 (519 queries consisted only of stop lemmas).

Average query times:

Idx1: 13.66 sec., Idx2: 0.29 sec.

Average data read sizes per query:

Idx1: 468.6 MB, Idx2: 10.236 MB.

We improved the query processing time by a factor of 47.1 with Idx2, and we improved the data read size by a factor of 45.77.

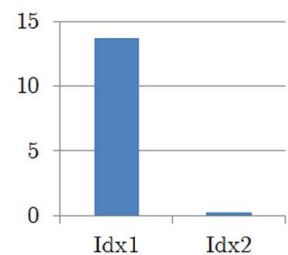

Fig. 5. Average query execution times for Idx1 and Idx2 (seconds), with MaxDistance = 9.

With *MaxDistance* = 9, we have the same average query execution time as with *MaxDistance* = 5. We can see a small increase in the average data read size per query in comparison with the *MaxDistance* = 5 case.

The average query execution times with the additional indexes are roughly the same with *MaxDistance* = 5, 7 and 9. The disposition of the data on the disk or some peculiarities of our index structure [13] could be sources of minor differences.

## XI. OTHER ADDITIONAL INDEXES AND RELATED WORK

In [6, 14, 15], nextword indexes and partial phrase indexes are introduced. These additional indexes can be used to improve performance. However, they can help only with phrase searches. Consider the text "to be or not to be". With the query "to be not to be", this text will not be found in a phrase search. Thus, our approach is more powerful.

Only phrase search is optimized in [16] as well.

In [1], only two-term queries are processed. The authors of [1] decreased the query processing time by up to a factor of 5 (table 5-2 in [1]). By contrast, our indexes can decrease the query processing time by up to a factor of 44-47, and we support multiple-term queries. We can see this in Fig. 6.

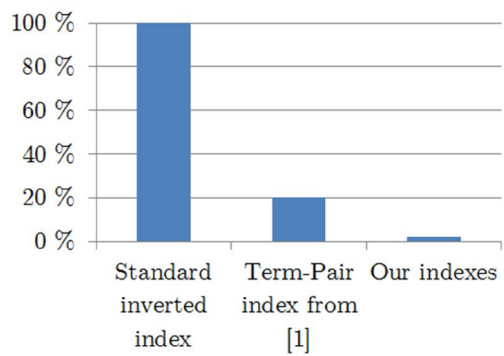

Fig. 6. Query processing time comparison with Term-Pair indexes [1].

The leftmost bar shows the average query execution time with the standard inverted indexes, normalized to 100%. The center bar shows the average execution time with term-pair indexes [1] relative to that with the standard inverted file. The rightmost bar shows the average execution time with our indexes relative to that with the standard inverted file. The rightmost bar is tiny because of our very fast searches.

## XII. VALUE OF MAXDISTANCE

The value of *MaxDistance* may be different for different types of lemmas. For example, for stop lemmas, we can use 5 or 7, whereas for frequently used lemmas, we can use 7, 9 or 11.

We can assume that for more frequently occurring lemmas, the importance of the semantic connections between nearby words will be high only for small distances between words. For less frequently occurring lemmas, the importance of semantic connections can be higher at larger distances.

Moreover, we can introduce a function *FMaxDistance*(*w*) to represent the value of *MaxDistance* for lemma *w*.

## XIII. CONCLUSION AND FUTURE WORK

In this paper, we have introduced several types of words and several types of additional indexes for different word types. We can use additional indexes of different types depending on the types of words contained in the search query.

A search query can contain any words, including very frequently occurring words.

We have also defined several types of search queries depending on the types of words they contain.

For each search query type, we have defined which types of additional indexes can be used for query execution.

We have presented the results of experiments showing that the average time of query execution with our indexes is 44-47 times less than that required when using ordinary inverted indexes.

For each word in the text, we use the additional indexes to store information about the words at distances from the given word of less than or equal to *MaxDistance* (a parameter, which can take a value of 5, 7, or even more). This information allows us to enhance the processing speed for frequently occurring words contained in the search query, such as "war", "world", "beautiful", "red", "mine", "be", and "who".

We also studied the dependence of the query execution time on the value of *MaxDistance*. The results of search experiments with *MaxDistance* = 5, 7, and 9 are presented.

In future research, we wish to study optimized methods of index creation for large values of *MaxDistance*. The index building time for large values (greater than 9) of *MaxDistance* can, for now, be regarded as a limitation of our method. Moreover, it will be interesting to investigate different types of queries in more detail.

REFERENCES

[1] H. Yan, S. Shi, F. Zhang, T. Suel, J. R. Wen, "Efficient term proximity search with term-pair indexes," CIKM '10 proceedings of the 19th ACM international conference on information and knowledge management. Toronto, ON, Canada, October 26–30, 2010, pp. 1229–1238.

[2] S. Buttcher, C. Clarke, B. Lushman, "Term proximity scoring for ad-hoc retrieval on very large text collections," In SIGIR'2006, pp. 621–622.

[3] J. Zobel, A. Moffat, "Inverted files for text search engines," ACM computing surveys, 2006, 38(2), Article 6.

[4] A. Tomasic, H. Garcia-Molina, K. Shoens, "Incremental updates of inverted lists for text document retrieval," SIGMOD '94 Proceedings of the 1994 ACM SIGMOD International Conference on Management of Data. Minneapolis, Minnesota, May 24–27, 1994, pp. 289–300.

[5] G. Zipf, "Relative frequency as a determinant of phonetic change," Harvard studies in classical philology. 1929, vol. 40, pp. 1–95.

[6] H. E. Williams, J. Zobel, D. Bahle, "Fast phrase querying with combined indexes," ACM transactions on information systems (TOIS). 2004, vol. 22, no. 4, pp. 573–594.

[7] R. Schenkel, A. Broschart, S. Hwang, M. Theobald, G. Weikum, "Efficient text proximity search," String processing and information retrieval. 14th International Symposium. SPIRE 2007. Lecture notes in computer science. Santiago de Chile, Chile, October 29–31, 2007. vol. 4726. Springer, Berlin, Heidelberg. pp. 287–299.

[8] S. Brin and L. Page. "The anatomy of a large-scale hypertextual web search engine," In Proc. of the 7th Intl. Conf. on World Wide Web (WWW'98), 1998.

[9] Veretennikov A.B. "O poiske fraz i naborov slov v polnotekstovom indekse [About phrases search in full-text index]," Sistemy upravleniya i informatsionnye tekhnologii [Control systems and information technologies]. 2012. vol. 48, no. 2.1, pp. 125–130. In Russian.

[10] Veretennikov A.B. "Ispol'zovanie dopolnitel'nykh indeksov dlya bolee bystrogo polnotekstovogo poiska fraz, vklyuchayushchikh chasto vstrechayushchiesya slova [Using additional indexes for fast full-text searching phrases that contains frequently used words]," Sistemy upravleniya i informatsionnye tekhnologii [Control Systems and Information Technologies]. 2013. vol. 52, no. 2, pp. 61–66. In Russian.

[11] Veretennikov A.B. "Effektivnyi polnotekstovyi poisk s ispol'zovaniem dopolnitel'nykh indeksov chasto vstrechayushchikhsya slov [Efficient full-text search by means of additional indexes of frequently used words]," Sistemy upravleniya i informatsionnye tekhnologii [Control Systems and Information Technologies]. 2016. vol. 66, no. 4, pp. 52–60. In Russian.

[12] Veretennikov A.B. "Sozdanie dopolnitel'nykh indeksov dlya bolee bystrogo polnotekstovogo poiska fraz, vklyuchayushchikh chasto vstrechayushchiesya slova [Creating additional indexes for fast full-text searching phrases that contains frequently used words]," Sistemy upravleniya i informatsionnye tekhnologii [Control systems and information technologies]. 2016. vol. 63. no. 1, pp. 27–33. In Russian.

[13] Veretennikov A.B. "O strukture legko obnovlyaemykh polnotekstovykh indeksov [About a structure of easy updatable full-text indexes], " Sovremennye problemy matematiki i ee prilozhenii. Trudy Mezhdunarodnoi (48-i Vserossiiskoi) molodezhnoi shkoly-konferentsii} [Proceedings of the 48th International Youth School-Conference "Modern Problems in Mathematics and its Applications"]. 2017. pp. 30–41. http://ceur-ws.org/Vol-1894/.

[14] D. Bahle, H. E. Williams, J. Zobel, "Efficient phrase querying with an auxiliary index," SIGIR '02 Proceedings of the 25th Annual International ACM SIGIR conference on research and development in information retrieval. Tampere, Finland, August 11–15, 2002, pp. 215–221.

[15] M. Chang, Chung Keung Poon, "Efficient phrase querying with common phrase index," ECIR 2006, LNCS 3936, Springer-Verlag Berlin Heidelberg, 2006, p. 61–71.

[16] Shashank Gugnani, Rajendra Kumar Roul, "Triple indexing: an efficient technique for fast phrase query evaluation," International journal of computer applications, 2014, vol.87, no 13, pp. 9–13.